\documentclass[preprint,11pt]{aastex}
\usepackage{epstopdf}
\begin{document}
\title{Synchrotron Emission on the Largest Scales: \\ 
                         Radio Detection of the Cosmic-Web} 
\author{Shea Brown\altaffilmark{1,2}} 
 
\altaffiltext{1}{CSIRO Astronomy \& Space Science; e-mail:shea.brown@csiro.au}
\altaffiltext{2}{Bolton Fellow}

\begin{abstract}Shocks and turbulence generated during large-scale structure formation are predicted to produce large-scale, low surface-brightness synchrotron emission. On the largest scales, this emission is globally correlated with the thermal baryon distribution, and constitutes the ``Synchrotron Cosmic- Web". I present the observational prospects and challenges for detecting this faint emission with upcoming SKA pathfinders.
\end{abstract}


\section{Introduction} In the current epoch roughly half of the number of baryons expected from concordance cosmology remain missing. At higher redshifts (z $>$ 2) these baryons are visible through  Ly$\alpha$ forest observations of the photoionized intergalactic medium (IGM) and ordinary galaxies (e.g. Rauch et al. 1997; Weinberg et al. 1997; Schaye 2001). Simulations suggest that the collapsing diffuse IGM was shock heated and now resides in filaments as T$\sim$10$^{5-7}$~K WHIM, where it is practically invisible at most wavelengths (Cen \& Ostriker 1999, 2006; Dave$^{\prime}$ et al. 2001). Absorption detections of the coolest WHIM components in the UV and X-ray have been reported (see Bregman et al. 2010, and references therein), but they give no information on the spatial distribution. However, shocks from infall into and along the filamentary structures between clusters are now widely expected to generate relativistic plasmas which track the distribution of the WHIM (Keshet, Waxman \& Loeb 2004; Pfrommer et al. 2006; Ryu et al. 2008; Skillman et al. 2008); Indeed, merger/accretion shocks are seen as polarized radio sources (the Ôperipheral relicsÕ) at the edges of the dense X-ray gas in clusters. This radio emission also has the potential for probing into the lower density regions further from cluster cores. When such features are detected, they can be used to set limits on the pressure of the (invisible) thermal gas, delineate shock structures and illuminate large-scale magnetic fields (Rudnick et al. 2009).

\section{The Synchrotron Cosmic-Web} 
\begin{figure}[h]
%\vspace*{-1mm}
\begin{center}  
\includegraphics[width=11cm]{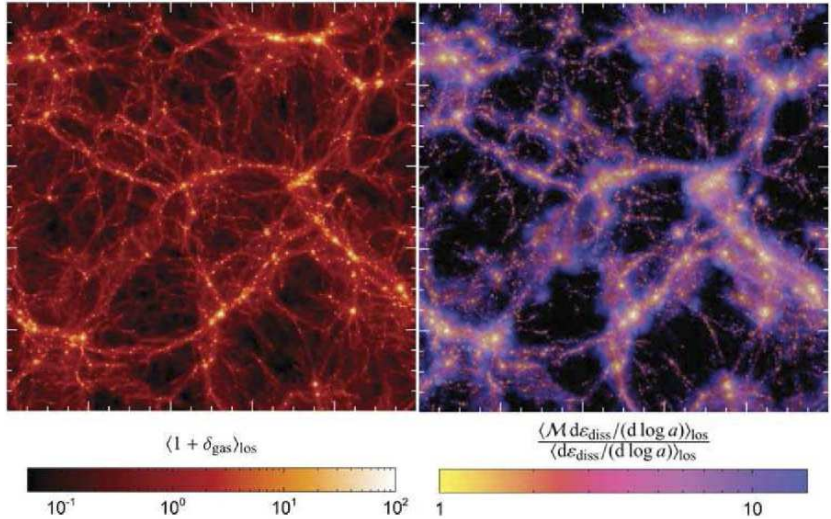}
\begin{footnotesize}
\caption{\label{web} Left: gas density from a cosmological simulation (Pfrommer et al. 2006); Right: Mach number of cosmic shocks, weighted by the energy injected into cosmic-rays. These two are correlated on galaxy cluster/filament scales.}
\end{footnotesize}
%\rule{17cm}{1pt}
\end{center}   
\end{figure}

The presence of faint synchrotron emission permeating large-scale structure (LSS), which we call the ``Synchrotron Cosmic-Web" (SCW), is a theoretical prediction that has yet to observationally confirmed (Wilcots 2004). Radio observations of some massive clusters of galaxies reveal large-scale synchrotron features, the so-called radio halos and peripheral relics, that give rise to the need for an in situ cosmic-ray (CR) acceleration process that has yet to be conclusively determined (see Brunetti this volume for a review). The physical mechanisms driving these candidate processes are shocks and turbulence in the IGM due to the gravitational buildup of LSS, which we predict from theory and numerical simulations should also be present in the lower density regions of the cosmic-web. Therefore, cosmological simulations that have been created to predict/test the radio synchrotron emission from massive clusters also predict low surface-brightness emission throughout LSS. Fig \ref{web} shows such a simulation (Pfrommer et al. 2006), which uses a diffusive shock-acceleration prescription for turning the energy in structure formation shocks into CR protons (CRp). It is expected that some fraction of the shock energy will also go into accelerating CR electrons (e.g., Drury 1983; Blandford \& Eichler 1987) which can emit synchrotron radiation in the presence of a cosmic magnetic field. However, inferring the presence of magnetic fields outside of clusters of galaxies is a difficult task (e.g., Xu et al. 2006), thus little is know observationally about the strength and ordering of magnetic fields in filaments.  Ryu et al. (2008) demonstrated that strong turbulence driven by structure formation shocks can generate current epoch (volume averaged) filament magnetic fields of the order of 0.01~$\mu$G, with a typical coherence length of a few $\times$~100~$h^{-1}$~kpc (Cho \& Ryu 2009). Mass weighted $\sqrt{<B^{2}>}$ values, which are appropriate for synchrotron emission, from the same simulation yields a characteristically higher value of 0.2~$\mu$G in filaments (Brown et al. 2010). Other simulations that consider cosmological fields seeded by galactic outflows, however, find only 0.005~$\mu$G current epoch magnetic fields in filaments (Donnert et al. 2009). Of the numerical simulations that produce synthetic radio maps, few focus on quantitative predictions for the SCW. Pfrommer et al. (2008) and  Battaglia et al. (2010) predict levels reaching at most a few $\mu$Jy~arcmin$^{-2}$ in the outskirts of massive clusters at 1.4~GHz. 

\section{Observational Obstacles and Issues}
\noindent {\sc Sensitivity: }\\ The predicted surface-brightnesses mentioned above, while extremely faint, are not out of reach for current radio telescope arrays. The raw thermal noise of a large array like the EVLA\footnote{https://science.nrao.edu/facilities/evla/} can reach these sensitivities in a reasonable amount of time. The real problem is the spatial filtering of large-scale emission that naturally arises with interferometric telescope arrays. The Fourier Transform nature of aperture synthesis means that the shortest baselines (correlation of the signal from the closest telescopes) are sensitive to the largest scale emission, but also the most sensitive to external/internal radio frequency interference (RFI). It is thus critical to obtain as many short baselines as possible, while insuring the least amount of corruption from RFI -- not an easy task. The nominal largest-scale that the EVLA can detect with a short integration at 1.4~GHz in its most compact configuration (D-Array) is roughly 15$^{\prime}$ (e.g., Rudnick \& Brown 2009), which is 1~Mpc at $z\sim0.06$. Thus, smooth emission associated with large-scale structure closer than $z\sim0.06$ will be resolved out by the EVLA. This can be improved by going to lower-frequencies and/or longer integration times (more short projected baselines as the source moves away from the zenith). 

\noindent {\sc Confusion:} \\ The most significant obstacle for reaching $\mu$Jy~arcmin$^{-2}$ sensitivities is confusion, both from resolved and un-resolved sources. The RMS confusion, $\sigma_{c}$, from unresolved continuum sources can be estimated by 

\begin{equation} \frac{\sigma_{c}}{mJy~beam^{-1}} \approx 0.2 \left(\frac{\nu}{GHz} \right)^{-0.7} \left( \frac{\theta}{arcmin} \right)^{2},
\end{equation}

\noindent where $\theta$ is the FWHM of the Gaussian beam. To overcome this limit, one must utilize sensitive observations at higher resolution in order to model and subtract the point-source contamination. Statistical reduction of this confusion is also possible (see below), as long as the sources causing the confusion are uncorrelated to large-scale structure. 

Ordinary galaxies, which produce synchrotron and free-free emission of their own, can also introduce significant confusion. More importantly, this confusion is perfectly correlated with LSS. Ponente et al. (2011) recently calculated the integrated surface-brightness of ordinary star-forming galaxies out to $z=10$ in an attempt to estimate their contribution to the cosmic radio background (CRB; Fixsen et al. 2010; Seiffert et al. 2010; Singal et al. 2010). The brightness temperature due to synchrotron radiation is 

\begin{equation} T_{syn}=0.1402~K~\left(\frac{\nu}{GHz}\right)^{-2.7}=~\frac{c^{2}I_{\nu}}{2k\nu^{2}}
\end{equation}

\noindent This translates into $I_{\nu}\sim$~0.4~mJy~arcmin$^{-2}$ at 1~GHz, which is well above the expected surface-brightness of filament fields. However, this includes the combined contribution from all normal galaxies out to $z=10$, and the integrated brightness for $z<0.3$ is an order of magnitude less. The brightness is less still when considering the emission due to a single redshift slice. In order to deal with this confusion, one needs to either work at a resolution that resolves the galaxy emission, or estimate and subtract the contamination through other means (e.g., the radio/IR correlation). In any case, this correlated emission sets the current fundamental limit for correlation studies (see below). 

\begin{figure}[h]
%\vspace*{-1mm}
\begin{center}  
\includegraphics[width=6cm]{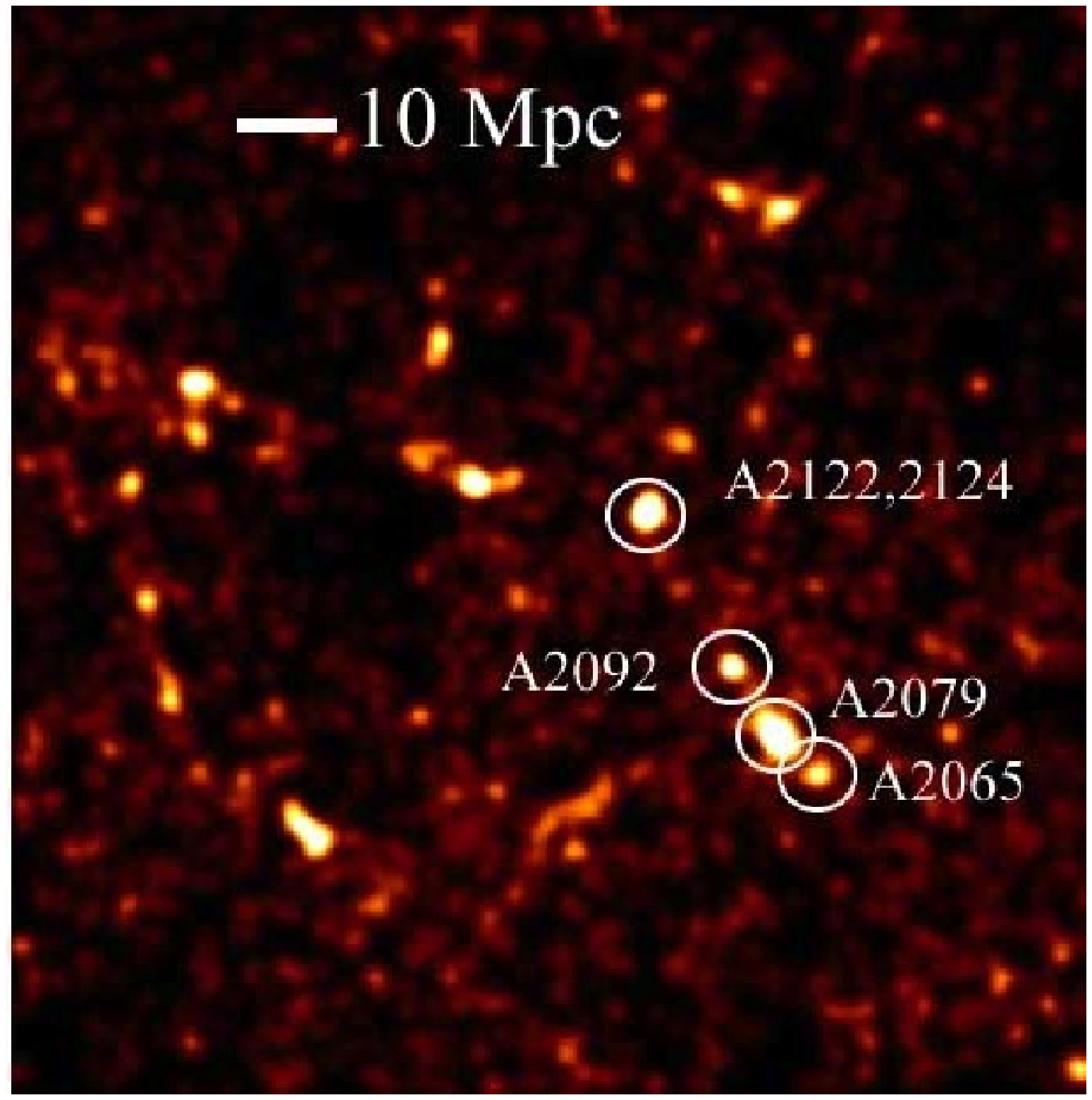}
\includegraphics[width=6cm]{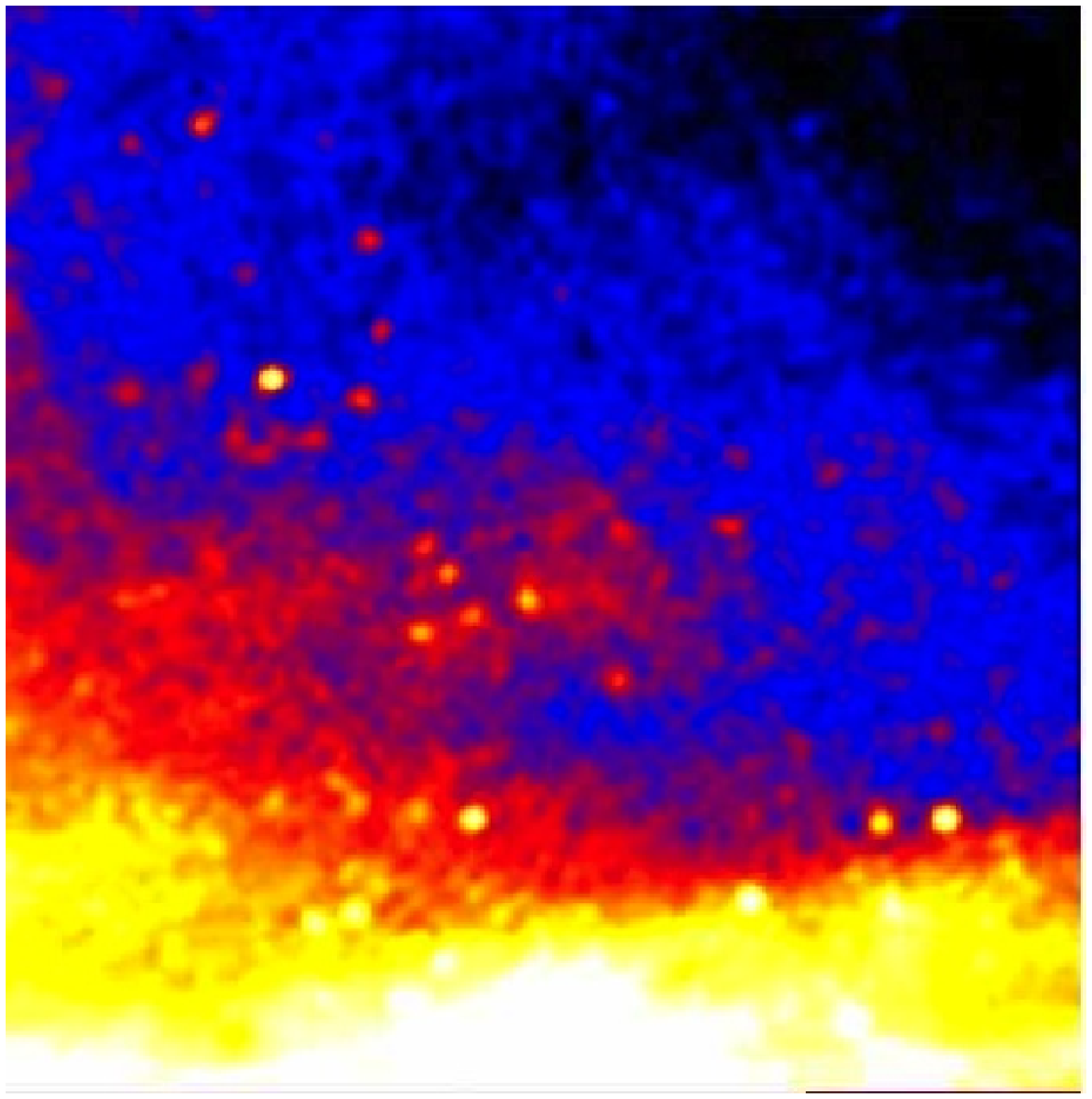}
\begin{footnotesize}
\caption{\label{bonn}Center coordinates for each image are 16h00m00s +35d00m00s (Brown et al. 2010).  Left: The surface-density of galaxies from 2MASS with 0.06 $<$ z $<$ 0.07, convolved to 36$^{\prime}$. Right: Radio continuum image of the same field at 1.4 GHz from the Stockert 25m telescope (36$^{\prime}$ resolution; Reich \& Reich 1986) }
\end{footnotesize}
\end{center}   
\end{figure}

Diffuse emission from the Milky Way presents yet another obstacle for the detection of the synchrotron cosmic-web. Synchrotron emission from the Milky Way dominates the SCW at all frequencies and scales, though surface-brightness fluctuations decrease at smaller angular scales (La Porta et al. 2008). Figure \ref{bonn} compares a map of large-scale structure (0.06 $<$ z $<$ 0.07) and the corresponding Galactic synchrotron emission from the Stockert 1.4~GHz survey (Reich \& Reich 1986). Most of the power from the Galaxy is on angular scales $\theta>$10~Mpc at these redshifts. Interferometric observations will filter most of this emission, though small-scale Galactic fluctuations are still an order of magnitude greater than the SCW. Unlike in CMB analysis, this foreground has a similar frequency behaviour to the SCW, and thus cannot be removed with template matching at other wavelengths (e.g., Gold et al. 2011). At the present time, statistical detection of the SCW may be the only option to get around the problem of confusion. \\

\noindent {\sc Statistical Methods:}\\ As is evident in Fig. 1, the SCW is spatially correlated with large-scale structure. Statistical methods can be used to infer the presence and average surface-brightness of this large-scale emission (e.g., Brown et al. 2010). Both stacking and cross-correlation have been used extensively in other areas of astrophysics and cosmology to detect faint, spatially correlated, emission below the noise limit of a given survey. These methods can be directly applied to the problem of detecting the SCW, though care must be taken to account for the sources of confusion mentioned above. 

The method of ``stacking" refers to the process of averaging together independent images of some class of objects, e.g, galaxies or clusters of galaxies, whose positions have been identified at other wavelengths. Even though the images are non-detections, by averaging $N$ images together, the signal-to-noise increases by $\sqrt{N}$, assuming the the source emission is spatially correlated and the image noise is truly independent. When searching for the SCW, it is not obvious what images to stack. The optical/IR distribution of galaxies can be used to trace large-scale structure, but the fact that these galaxies do not perfectly trace the diffuse thermal baryon distribution, and that the SCW is in turn not perfectly correlated with the diffuse thermal baryon distribution, means that the signal will not add perfectly coherently. This in turn means that the signal-to-noise will not increase as $\sqrt{N}$. The presence of radio emission from the galaxies making up the LSS (see above) means that some of the confusion noise will add coherently as well, further reducing the effectiveness of stacking. The most effective method of stacking LSS in order to detect the SCW remains to be determined, though the next generation of all-sky radio surveys (see below) may provide sufficient sky coverage and sensitivity to make stacking a promising technique. 

\begin{figure}[h]
%\vspace*{-1mm}
\begin{center}  
\includegraphics[width=6cm]{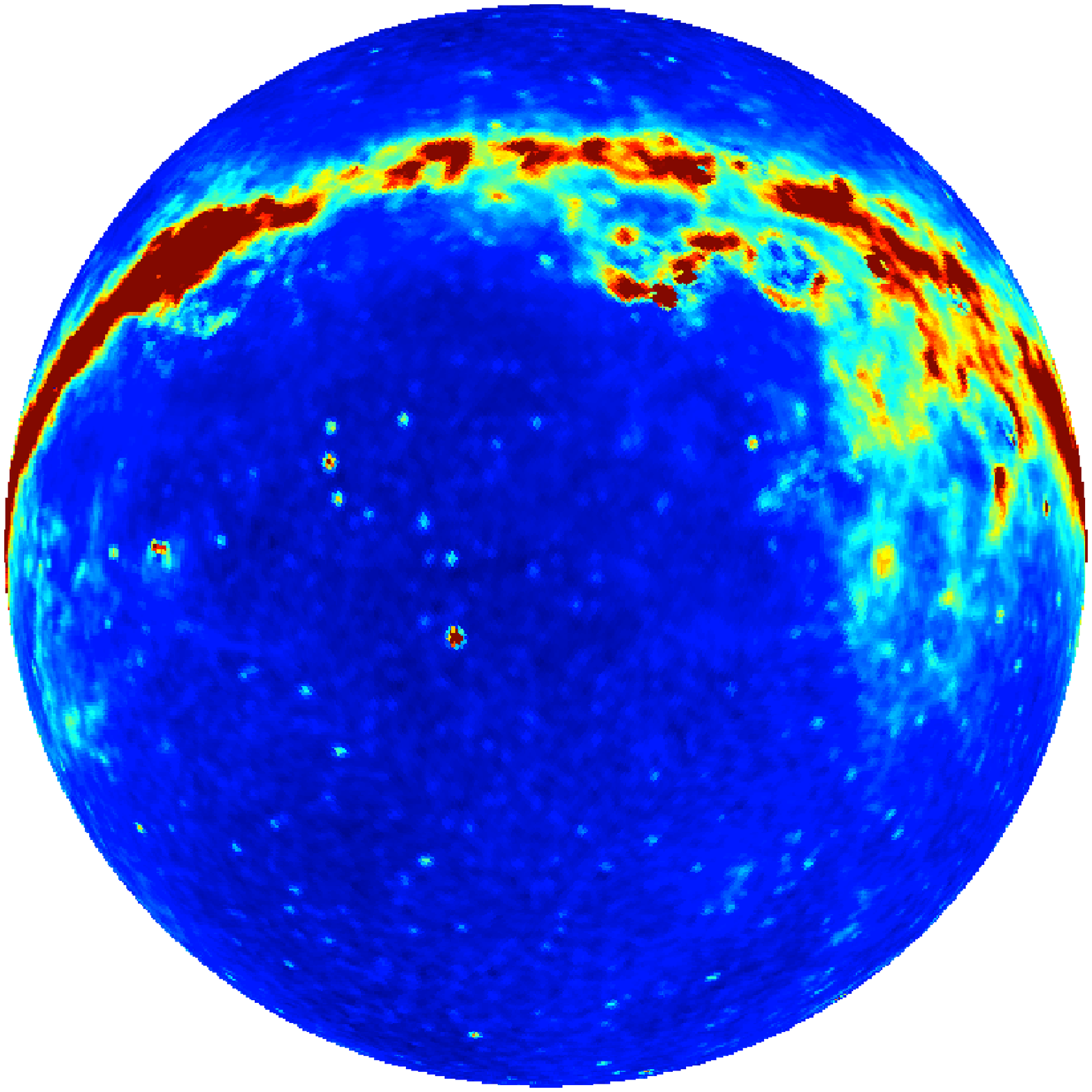}
\includegraphics[width=6cm]{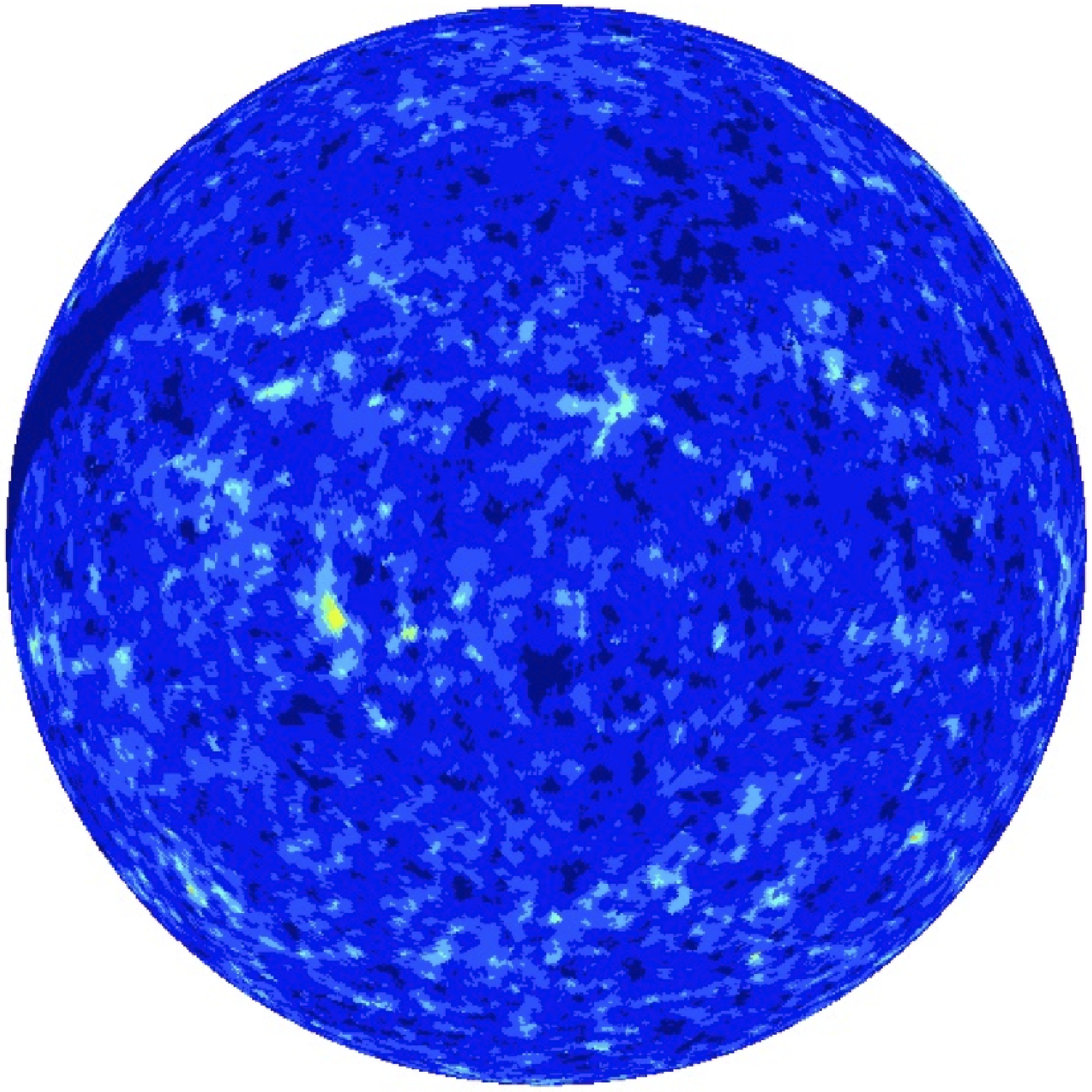}
\begin{footnotesize}
\caption{\label{cc} Left: WMAP 7yr K-band synchrotron map (image $\mathcal{R}$, see text); Right: 2MASS surface density of galaxies (13 $<$ K $<$ 13.5; image $\mathcal{G}$). The goal of cross-correlation is to detect evidence of our model for the SCW, which is a function of $\mathcal{G}$, in map $\mathcal{R}$. }
\end{footnotesize}
%\rule{17cm}{1pt}
\end{center}   
\end{figure}

Cross-correlation is another technique to statistically detect emission below the noise level of a map. The widest use of this technique has been in the CMB community, where it has been used to detect the late Integrated Sachs-Wolfe (ISW) effect (e.g., Crittenden \& Turok 1996; Boughn \& Crittenden 2004; Nolta et al. 2004; McEwen et al. 2007; Liu et al. 2011). The task of cross-correlation is to assess the extent to which image $\mathcal{R}$ (e.g., an all-sky radio map of synchrotron emission; Fig \ref{cc} left) is correlated with image $\mathcal{G}$ (e.g., a map of the distribution of galaxies at some redshift; Fig. \ref{cc} right). However, what one wants to do is assess the probability that a signal $m$, represented by a model that is some function of image $\mathcal{G}$, is present in image $\mathcal{R}$. Cross-correlation can be performed in many ways, and has been attempted successfully in real (e.g., Nolta et al. 2004), harmonic (Afshordi et al. 2004), and wavelet (McEwen et al. 2007) space. The benefit of cross-correlation, relative to stacking, is that it makes use of all the data available to assess the probability. However, interpreting the results of a cross-correlation experiment is more difficult, as one must account for all possible sources of correlation. For example, Fig. \ref{cc} shows that, on large scales, the surface-density of galaxies (right) is {\it anti}correlated with the synchrotron brightness (left), because the latter also traces areas of high optical depth through the Milky Way. It is this effect that gives rise to need to mask low Galactic latitude regions of both images before cross-correlation, as is typically done in CMB analysis. The presence of correlated confusion mentioned above means that high resolution radio images are needed in order to make cross-correlation a feasible technique for detecting the SCW. 

\section{The Future} A number of pathfinders for the Square Kilometre Array (SKA) are being built and tested around the world. ASKAP\footnote{http://www.atnf.csiro.au/SKA/}, EVLA\footnote{https://science.nrao.edu/facilities/evla}, LOFAR\footnote{http://www.lofar.org/}, LWA\footnote{http://www.phys.unm.edu/~lwa/index.html}, Meerkat\footnote{http://www.ska.ac.za/meerkat/}, MWA\footnote{http://www.mwatelescope.org/}, and more. These telescopes will push the parameter-space of radio observations in the areas of sensitivity, sky coverage, and frequency. Below, I will briefly address how each of these frontiers can be used to probe the SCW, and which telescopes are relevant for these advances. \\

\noindent {\sc Deeper:} \\ Deeper observations with higher resolution of individual cluster outskirts and filaments are needed for direct detection of the synchrotron cosmic-web. Targeted observations of the peripheries of massive clusters and super-clusters is the most promising method for direct detection of the SCW, as these are the locations of the most powerful shocks and turbulence during LSS formation. The EVLA, LOFAR with its international baselines (eLOFAR), and eventually Meerkat, will be ideal for these observations. The removal of confusion using the long baselines, as mentioned above, will still be a big obstacle, and it has not been demonstrated that this can be done to reach $\mu$Jy~arcmin$^{-2}$ surface-brightness sensitivity yet. \\

\noindent {\sc Wider:} \\ Planned all-sky surveys with upcoming SKA pathfinders will be commencing in the coming years. These surveys will allow statistical detection through cross-correlation and stacking, subject to the constraints outlined above. The deepest of these surveys may also be able to directly detect the top end of the ``cosmic-shock" luminosity function (Battaglia et al. 2010; Skillman et al. 2010), which may provide candidate sites for deeper follow-up observations. The Evolutionary Map of the Universe (EMU)\footnote{http://www.atnf.csiro.au/people/rnorris/emu/} survey with ASKAP (Norris et al. this volume), as well as the Surveys Key Science Project with LOFAR (Heald et al. this volume) are two such surveys. \\

\noindent {\sc Lower Frequency:} \\ The steep spectral indices of expected in the SCW ($I_{\nu}\sim \nu^{-\alpha}$, where $\alpha > 1$) relative to both point and diffuse sources of confusion means that, if resolution is not sacrificed too much, going to lower frequencies is ideal for detecting sychrotron emission in LSS. Telescopes such as eLOFAR, the LWA, a possible low frequency upgrade to the EVLA (Kassim 2010), and the combination of MWA data with high resolution Giant Metrewave Radio Telescope (GMRT) observations, will operate in this regime.  Working at the lowest frequencies, however, will likely result in a sacrifice of polarization information (e.g., Farnsworth et al. 2011). \\

\noindent {\sc Higher Frequency:} \\ At frequencies greater then 1~GHz, radio observations will benefit from higher resolution, but suffer from lower intrinsic source surface-brightness. Polarization observations will provide the biggest gains, especially when searching for the shock-structures themselves. Structure formation shocks onto and along filaments are predicted to be narrow, flat spectrum ($\alpha < $1), and highly polarized (e.g., Ryu et al.; Skillman et al. 2010). Therefore, both direct and statistical detection of these shocks can ideally be performed at $\nu >$1~GHz.  The Polarization Sky Survey of the Universe's Magnetism (POSSUM; Gaensler et al. 2010) with ASKAP, which will produce polarized maps of the southern sky from 1130-1430~MHz, is one such survey that has the potential to detect these polarized cosmic-shocks.  

\section{References}

\noindent Afshordi, N., Loh, Y.-S., \& Strauss, M.~A.\ 2004, \prd, 69, 083524 \\
 
\noindent Battaglia, N., 
Pfrommer, C., Sievers, J.~L., Bond, J.~R., 
\& En{\ss}lin, T.~A.\ 2009, \mnras, 393, 1073  \\
 
 \noindent Brown, S., Farnsworth, 
D., \& Rudnick, L.\ 2010, \mnras, 402, 2 \\

\noindent Boughn, S., \& Crittenden, R.\ 2004, \nat, 427, 45 \\
 
\noindent Bregman, J.~N., et al.\ 2009, astro2010: The Astronomy and Astrophysics Decadal Survey, 2010, 24 \\

\noindent Cen, R., \& Ostriker, J.~P.\ 1999, \apj, 514, 1 \\

  \noindent Crittenden, R.~G., \& Turok, N.\ 1996, Physical Review Letters, 76, 575 \\
  
\noindent Dav{\'e}, R., et al.\ 
2001, \apj, 552, 473 \\

  \noindent Donnert, J., Dolag, K., 
Lesch, H., Muuller, E.\ 2009, \mnras, 392, 1008 \\

\noindent Farnsworth, D., 
Rudnick, L., \& Brown, S.\ 2011, \aj, 141, 191 \\
 
 \noindent Gaensler, B.~M., Landecker, T.~L., Taylor, A.~R., \& POSSUM Collaboration 2010, Bulletin of the American Astronomical Society, 42, \#470.13 \\
  
 \noindent Gold, B., et al.\ 2011, \apjs, 192, 15 \\

 \noindent Kassim, N.\ 2010, Galaxy Clusters: Observations, Physics and Cosmology, 12P \\
  
  \noindent Keshet, U., Waxman, E., 
\& Loeb, A.\ 2004, \apj, 617, 281 \\
  
  \noindent La Porta, L., Burigana, C., Reich, W., \& Reich, P.\ 2008, \aap, 479, 641 \\

 \noindent Liu, G.-C., Ng, K.-W., \& Pen, U.-L.\ 2011, \prd, 83, 063001 \\

\noindent McEwen, J.~D., Vielva, 
P., Hobson, M.~P., Mart{\'{\i}}nez-Gonz{\'a}lez, E., 
\& Lasenby, A.~N.\ 2007, \mnras, 376, 1211 \\

  \noindent Nolta, M.~R., et al.\ 2004, \apj, 608, 10 \\
  
  \noindent Pfrommer, C., 
Springel, V., En{\ss}lin, T.~A., \& Jubelgas, M.\ 2006, \mnras, 367, 113 \\ 

\noindent Pfrommer, C., 
En{\ss}lin, T.~A., \& Springel, V.\ 2008, \mnras, 385, 1211 \\

  \noindent Rauch, M., et al.\ 1997, 
\apj, 489, 7 \\

  \noindent Reich, P., \& Reich, W.\ 1986, \aaps, 63, 205 \\

  \noindent Rudnick, L., \& Brown, S.\ 2009, \aj, 137, 145 \\
  
  \noindent Rudnick, L., et al.\ 
2009, astro2010: The Astronomy and Astrophysics Decadal Survey, 2010, 253 \\

  \noindent Ryu, D., Kang, H., Hallman, 
E., \& Jones, T.~W.\ 2003, \apj, 593, 599 \\
  
  \noindent Ryu, D., Kang, H., Cho, J., 
\& Das, S.\ 2008, Science, 320, 909 \\

\noindent Schaye, J.\ 2001, \apjl, 562, 
L95 \\

\noindent Skillman, S.~W., 
O'Shea, B.~W., Hallman, E.~J., Burns, J.~O., 
\& Norman, M.~L.\ 2008, \apj, 689, 1063 \\

\noindent Weinberg, D.~H., 
Miralda-Escude, J., Hernquist, L., \& Katz, N.\ 1997, \apj, 490, 564 \\

\noindent Wilcots, E.\ 2004, \nar, 48, 1281\\

\label{lastpage}
\end{document}